# Generation and escape of local waves from the boundary of uncoupled cardiac tissue

*Vadim N.Biktashev*, Ara Arutunyan[#], Narine A. Sarvazyan[#]*

*Department of Mathematical Sciences, University of Liverpool, UK
[#]Pharmacology and Physiology Department, The George Washington University,
Washington DC, USA

Corresponding author:
Narine Sarvazyan, Ph.D.
Pharmacology and Physiology Department
The George Washington University
2300 Eye Street, Washington DC 20037
Phone (202)994-0626, Fax: (202)994-3553,
Email: phynas@gwumc.edu

Abstract

We aim to understand the formation of abnormal waves of activity from myocardial regions with diminished cell-to-cell coupling. In route to this goal, we studied the behavior of a heterogeneous myocyte network in which a sharp coupling gradient was placed under conditions of increasing network automaticity. Experiments were conducted in monolayers of neonatal rat cardiomyocytes using heptanol and isoproterenol as means of altering cell-to-cell coupling and automaticity respectively. Experimental findings were explained and expanded using a modified Beeler-Reuter numerical model. The data suggests that the combination of a heterogeneous substrate, a gradient of coupling and an increase in oscillatory activity of individual cells creates a rich set of behaviors associated with self-generated spiral waves and ectopic sources. Spiral waves feature a flattened shape and a pin-unpin drift type of tip motion. These intercellular waves are action-potential based and can be visualized with either voltage or calcium transient measurements. A source/load mismatch on the interface between the boundary and well-coupled layers can lock wavefronts emanating from both ectopic sources and rotating waves within the inner layers of the coupling gradient. A numerical approach allowed us to explore how: i) the spatial distribution of cells, ii) the amplitude and dispersion of cell automaticity, iii) and the speed at which the coupling gradient moves in space, affects wave behavior, including its escape into well-coupled tissue.

INTRODUCTION

A bulk of evidence suggests that arrhythmogenic ectopic beats may originate from the areas of diminished cell-to-cell coupling (1). Such areas can be anatomical or functional. Anatomical examples of myocardial tissue with diminished coupling include an infarct scar (2,3), inflammatory infiltration (4), diffuse fibrofatty tissue responsible for arrhythmogenic right ventricular dysplasia syndrome (5), changes in myocardial fiber orientation (6), or an island of engrafted stem cells (7). Functional uncoupling occurs during ischemia, due to the acidic environment and fatty acid accumulation, both of which diminish gap junctional conductance (1,8,9).

Between myocytes within areas of diminished cell-to-cell coupling and the surrounding, well-coupled cell layers, there is a boundary layer with transitional values of coupling. This study considers possible behavior of tissue within such a boundary layer during conditions that promote cell automaticity. The latter can occur during ischemia or reperfusion as a result of catecholamine release or calcium overload, respectively. Indeed, it has been shown that ischemia leads to a hundred-fold increase in concentration of interstitial catecholamines which comes from ischemic nerve endings (10). Thus, it is likely that during conditions of diminished blood flow the layers of poorly coupled cells can be awash in norepinephrine-containing interstitial fluid. The effects could be further exaggerated in areas which exhibit elevated adrenergic responsiveness, an effect known as denervation supersensitivity (11). Reperfusion-associated automaticity, on other hand, is attributed to calcium-overload (12). Diffusion of neurotransmitters from neighboring regions contributes to the reperfusion-induced increase in intracellular calcium, giving rise to triggered activity (13).

The boundary layer is likely to comprise a small volume of tissue (few mm wide, or few hundreds of cells across) and be hidden under layers of normally oxygenated myocardium. Thus, on a whole heart level, an abnormal wave emanating from such a region can appear as a single ectopic beat. To date, little is known of what lies beneath these macroscopic events. This is because the technical means to visualize the initial steps of ectopic beat generation in vivo, on a cellular level, are yet to be developed. Therefore we attempted to gain initial insights into this process using *in vitro* networks of cardiac cells and to expand our experimental findings with numerical studies. Specifically, we asked: what is the behavior of a heterogeneous cardiac cell network when a gradient of cell-to-cell coupling is superimposed with an increase in cell automaticity? The experimental and modeling data suggested the existence of a rich and interesting behavior which included the formation of multiple automatic sources and spiral waves. Their behavior and ultimate fate were dependent on the movement of the boundary in space, degree of cell network heterogeneity, and other factors considered below.

METHODS

Cardiomyocyte culture. Cardiomyocytes from two-day old Sprague-Dawley rats were obtained using an enzymatic digestion procedure (14) in accordance with the guidelines of the Institutional Animal Care and Use Committee. The cells were plated on 25-mm laminin-coated glass coverslips ($10^5$ cells/cm$^2$) and kept under standard culture conditions in Dulbecco-modified minimum essential medium supplemented with 5% FBS, 10 U/ml penicillin, 10 µg/ml gentamicin and 1 µg/ml streptomycin. By the third day in culture, the cells had formed interconnected confluent networks and were used in experiments for an additional 3-4 days.

Experimental chamber. A custom-made experimental chamber was used to perfuse a small area of a cell network with a solution of interest, while observing events under the microscope (Fig.1A). The design of the chamber and its flow characteristics have been previously described (14,15). It uses a stainless steel holder to mount a glass coverslip on the raised surface of a plastic holder, which contains two inlets and one outlet (Fig.1). The polished sides of the chamber provide an airtight contact with the coverslip, whereas the Plexiglas ceiling creates a 300 μm perfusion space. Superfusion solutions are driven by a multisyringe pump (Harvard Apparatus) loaded with 10 and 30 ml glass syringes.

Experimental protocol. First an island of uncoupled cells was created by locally applying 2 mM heptanol (Fig.1B, area shown in solid gray). In the past we have shown that 2mM heptanol fully uncouples cells in our preparations (16,17). Washout was started by switching OFF inlet #2 and allowing 5 μmol/L isoproterenol-containing Tyrode to gradually shrink the uncoupler-containing inner area. Because chamber has a closed design (Fig. 1A, right), such shrinkage occurs only from the sides as illustrated in Fig.1B. Cells within the slightly shaded area are partially uncoupled, not because of diluted heptanol, but because of the time for this cells to recover from the effects of the uncoupler. Myocytes on the boundary between the two regions experienced the concurrent changes in coupling and automaticity as cells recovered from the uncoupler and cAMP-mediated effects of isoproterenol were taking place. The washout process lasted 2.5 min, after which the area where the waves did not propagate disappeared.

Monitoring network behaviour. Cells plated on laminin-covered coverslips were loaded with 5 μM Fluo-4AM for 1 h. Each spontaneous or paced action potential was associated with a calcium transient. Fluo-4 was excited at 488 nm, and the fluorescence was acquired at wavelengths of >515 nm. Experiments were conducted using a BioRad MRC-1024 confocal imaging system using low power magnification objective (Olympus PlanApo 4X/0.16NA). Conclusions are based on 22 experiments using 5 different cell preparations. The number of recorded events for a specific scenario is mentioned in parenthesis within the corresponding sentence. Notably, the recorded cases are only a subset of the visually observed scenarios, i.e., the described sequence of events was observed in more experiments than it was recorded.

Numerical model. We used a generic Beeler-Reuter model of a cardiac myocyte (18), which contains an explicit, albeit simplified description of individual ionic currents. In the past this model has proven to be an adequate tool to closely describe the events seen in our experimental preparations (17). The advantage of the model is that it is computationally simple, allowing us to simulate hundreds of thousands of cells with relative ease. It also offers a convenient way to modulate cell automaticity (details below).

The state of a cell is described by the membrane potential $V$ satisfying

$$\partial_t V = -(I_{Na} + I_s + I_{K1} + I_{x1})/C + \text{coupling term}$$

where $C$ is the capacitance per area of membrane, $I_{Na}$ is the fast depolarizing sodium current, $I_s$ is the slow depolarizing current, carried mostly by calcium, $I_{K1}$ and $I_{x1}$ are two repolarizing potassium currents. These currents depend on membrane potential, cytosolic calcium concentration, and six gating variables. The currents were modified from the original Beeler-Reuter model: the gated ("time-dependent") component of $I_{Na}$ was 60% of its standard value, $I_s$ was 50% of its standard value (19), and $I_{K1}$ was modified in a complex time- and space-dependent way as explained within the text. The voltage-dependent functions were tabulated for $V$ in the range from −100 mV to 80 mV with a step 0.1 mV. The time-stepping was done using an explicit Euler scheme for all variables with a timestep of 0.1 ms. In simulations which employed a wider range of coupling values (example shown in Fig.5) the timestep was decreased to 0.02 ms for stability purposes. The choice of voltage and time steps was dictated by consideration of accuracy and stability, and was verified by varying the timestep and to ensure that no essential changes in the features of interest occurred. The spatial step corresponded to typical intercellular distance and was not varied.

Spatial arrangement of cells in numerical model
We considered an idealized situation, where cells are located on a square lattice, so the variables are labeled by two integers, $i$ and $j$, which label the rows and columns of the lattice (Fig.2, step 1). Cells are coupled to their nearest neighbors,

$$\text{coupling term}(i; j) = D/l^2 \, [V_{i+1; j}+V_{i-1; j}+V_{i; j+1}+V_{i; j-1} - 4 \, V_{i; j}] \qquad (*)$$

where the effective diffusion coefficient D, is proportional to the conductivity between cells, and $l$ is the distance between cells. We stress that equation (*) is not thought of here as a spatial discreziation of a Laplacian term as in a "reaction-diffusion" system, but represents the Ohm's and Kirchhoff's laws and the assumed rectangular geometry of the grid of individual cells. Currents through boundaries of the grid were assumed to be zero. The value of $l$ is set at 30 μm to account for the mean spacing between the centers of two adjacent cells, an estimate from experimental preparations. Velocity of propagation in isotropic cardiomyocyte networks at room temperature is ~10-15 cm/sec, which corresponds to $D \sim 0.10$ cm$^2$/sec (17,20).

The gradient of coupling strength was oriented vertically as shown in Fig.2, step 1, with the y axis running from the bottom to the top. The upper layers corresponded to a more coupled region ($D_{max}$), while the bottom ones to a fully uncoupled region ($D_{min}$). An exponential gradient between the two D values was then applied to the middle, or what we will call hereafter the "boundary layer".

Numerical means to increase cell automaticity
Our aim was to describe a heterogeneous network which becomes spontaneously active as excitatory effects of isoproterenol or barium are developed. Thus, to make cells spontaneously active we altered the balance between inward and outward currents by inhibiting the inward potassium rectifier current, $I_{K1}$, an approach taken by us and others in the past (17,21,22). By setting the initial values of $g_{K_1}$ at 30%, we were able to mimic the smaller $I_{K1}$ contribution reported for neonatal cardiomyocytes (23) (24) as compared to the original Beeler and Reuter values for adult ventricular cells (18). Further decrease in the channel conductance $g_{K_1}$ led to a spontaneous firing of individual cells as it alters the balance between inward and outward currents (17). Changes to the conductance were implemented as $g_{K_1}=0.3 - α(x,y,t)$, where parameter $α$, hereafter referred to as "automaticity", was varied in space and time, with larger alpha values corresponding to a higher automaticity. For the bottom, fully uncoupled cells, α values were set to zero to reflect the fact that in our experiments neither isoproterenol nor barium was present in the heptanol solution (Fig.2, steps 2&3). Notably, having or not having high automaticity in the inner zone would not make a difference in network behavior because cells there are fully uncoupled.

We want to stress that in contrast to the spatial gradient of cell-to-cell coupling (as detailed below), α values exhibited step-change across few cells on the lower border. This was done to mimic our experimental setup, in which mixing/diffusion between the two flows creates a physical gradient of about 180 microns between the two solutions (14).

Numerical means to implement cell heterogeneity

The heterogeneity between individual cells was implemented by introducing two coefficients: $\eta(x, y)$ and $\delta$. The first coefficient, $\eta(x, y)$ is a Gaussian distributed uncorrelated random variable implemented using the Box-Muller 1958 transformation with a mean of zero and standard deviation of 1. It allowed us to randomly distribute cells with different properties in space. The second coefficient, $\delta$, was introduced to describe the degree of dispersion from the mean value of automaticity $<\alpha>$. For an individual cell with coordinates $(x,y)$ automaticity was therefore described as $\alpha(x, y) = <\alpha> (1 + \delta \eta(x, y)$. Temporal changes of automaticity were implemented by changing the $<\alpha>$ value accordingly (summary in Table 1 and Figure legends). These changes were implemented for the boundary and the upper layer. For the bottom, uncoupled cells, α values remained set to zero (Fig.2, step 3). For the reader's convenience we compiled all the values used in selected simulations and their corresponding video files in Table 1.

Boundary movement in space. Our experimental settings mimicked a situation in which the ischemic (i.e., uncoupling) environment moves in space. In vivo, such a movement can be caused by reperfusion or by blood flow from a neighboring coronary bed as a result of local hyperemia. The boundary movement was implemented numerically as shown in Fig.2, step 4. The mean automaticity values were set to a fixed value associated with multiple local waves of activity (see Table 1 for the parameters used in the specific simulations) and the coupling gradient was moved downward with a constant speed. The automaticity values moved together with the coupling gradient as shown in step 4 of Fig.2. The speed of the downward movement was varied around the values seen in experiments.

RESULTS

PART I. Experimental studies

Observing the boundary on a cellular scale. Cells within the boundary layer between the two regions (denoted as a gradient of gray color, Fig.1B) experienced concurrent changes in coupling and automaticity. These changes occurred as a result of i) recovery from the heptanol and ii) cyclic AMP-mediated effects of isoproterenol (16). These conditions led to a formation of multiple ectopic sources (Fig.3), which appeared as multiple or individual waves next to the shrinking boundary (number of recorded cases: n =11). Their wavefronts fused and spread outwards giving an impression of a single ectopic source (Fig.3A top row). The position of the acquisition window relative to the uncoupled area is shown on the left (black box). It closely follows the spiral wave movement and therefore travels along the shrinking boundary. The speed at which experimental boundary moved in space was not the same at all points along the boundary (due to the geometry of the inner area and the way it shrinks upon washout), however, it was estimated to be in the range of 10-50 μm/sec, or approximately 1/2 - 2 cell/sec.

Spiral waves along the boundary. Ectopic sources gave rise to spiral waves, with durations ranging from 2 to 30 sec (number of recorded cases: n = 6). Both clockwise and counterclockwise spiral waves were observed (Fig.4A, Fig.3A, middle row and the corresponding video files). Once a spiral fully developed, it swept the entire cell network except the inner area blocked by the uncoupler. This suppressed individual ectopic sources, but they reappeared immediately after the spiral wave died off (number of recorded cases: n =5). Another common feature observed in all cases was flattening of the spiral shape. This effect is due to a steep gradient of coupling, and the corresponding gradient of conduction velocity: faster in the outside area, slower within the boundary. This sharp coupling gradient was also responsible for a wave "shedding" effect: near the boundary the wavelength becomes so short that wavefronts from two or more previous turns of a spiral were seen simultaneously next to each other (Fig.3B and the corresponding video file). The white arrows point to the wavefronts from two sequential turns of the same spiral.

Trajectory of the spiral tip: sequential pinning. The majority of spirals born within the boundary layer represented transient events lasting 2-5 rotations. If a spiral persisted longer than a few seconds it tended to travel along the interface between coupled and uncoupled cells (number of recorded cases: n=6). As it did, the motion of the spiral tip was noticeably non-stationary, with segments of pinned behavior alternating with periods of shifts along the boundary (Fig.4A and the corresponding video file). With the orientation of the coupling gradient employed here, counter-clockwise rotating spirals tended to drift rightwards, and the clockwise rotating spirals drifted leftwards. This is shown in Fig. 4A via tip trajectories and can be observed during the first 6 sec of the video corresponding to Fig.3 (leftward drift of a clockwise spiral – first 6 sec of the video; rightward drift of a counterclockwise spiral - 18-21s). In several cases (n=4), spirals appeared to be anchored to the areas, which moments later emitted circular ectopic waves (Figs.3A and 4B). This suggests pinning of the spiral waves to the areas of altered automaticity, which agrees with our numerical studies discussed below.

Alternative treatments and control experiments. We tested an alternative approach to elevate cell automaticity. Specifically, in the above experiments we substituted isoproterenol with barium chloride. Barium has been used as a tool to elevate cell automaticity by us and others (17,25). It does so by its direct inhibitory effect on inward potassium rectifier current. Similar to the isoproterenol experiments described above, barium led to multiple localized rotating waves and ectopic sources on the boundary of the heptanol-containing area.

Control experiments were conducted using one treatment agent. Specifically, when isoproterenol was omitted from the perfusate, no ectopic activity was observed and the heptanol-containing area shrunk without apparent impact on the rest of the cell layer (n=8). Effect of isoproterenol application without heptanol was associated with an increase in a monolayer's endogenous spontaneous firing rate (n=4). The result was a single uniform wave rapidly passing thru the entire coverslip (16). Barium application had a similar effect and was quantified by us previously (17). All-in-all, when applied by themselves, neither heptanol, isoproterenol nor barium produced patterns associated with local waves.

We note that macroscopic spiral waves can be readily induced in cardiomyocyte monolayers by either rapid pacing, cross-field or premature stimulation (26-28). We stress that the spiral activity reported here was not induced by external electrodes. The small rotating waves occurred *spontaneously* as a result of rapidly changing conditions on the boundary. Our next step was to explore these patterns numerically, as reported below.

PART II. Numerical studies
Behavior of the boundary layer. We started by creating a steep gradient in coupling (Fig.2, step 1), represented by the coefficient *D*, stretching from what is considered to be normal *D* values ($10^{-1}$ cm$^2$/sec), to a fully uncoupled cell network ($D=10^{-5}$ cm²/sec). Notably, the fourth order magnitude change in D values was not an arbitrary choice, but was dictated by experimentally observed propagation velocities. The latter differed by two orders of magnitude, from 12 cm/sec in fully coupled cultures (17) to ~0.1 cm/sec near the uncoupled area (see Fig. 3B: the scale bar is 0.5mm and the time stamp is on the top of the figures). The corresponding estimates of the gap junction conductivities range between 300 and 0.03 nS, which matches experimental and numerical data by others (see (29), figs 16 and 18D). One should keep in mind that propagation at higher automaticity is akin to phase waves and requires much smaller conductivity than normal excitation waves, and our simulations, as well as experiments, spanned parametric areas where the propagation in fact did not happen.

Mean cell automaticity, represented by value $<\alpha>$, was set to increase within the boundary and the upper regions at a steady rate of $d<\alpha>/dt = 0.001$ sec$^{-1}$ (Fig.2, steps 2&3). The choice of $<\alpha>$ values was not arbitrary, but spanned values at which the network was quiescent to the values at which local waves appeared (17). Importantly, this increase in automaticity occurred non-uniformly, depending on individual cells' $\alpha$ values, with dispersion coefficient $\delta = 0.5$. Graphically, mean automaticity $<\alpha>$ is shown as a bold grey line on the left of Fig.2, step 2, while a thin grey line shows the distribution of individual $\alpha$ values for a column of cells with a fixed $x$ coordinate.

Let us consider an example of one of these studies (Fig.5). The specific parameters for each figure can be found in Table 1. The $<\alpha>$ starts at 0.08 and increases with a rate of 0.001 sec$^{-1}$. For the first 25 sec no activity is observed in any of the zones (Fig.5A, first panel). At about $t = 27$ sec ($<\alpha>=0.107$), several small ectopic sources start to appear at the cell layer with $D \sim 10^{-3}$ cm$^2$/sec (Fig.5A, second panel). Waves generated by these ectopic sources do not spread to the upper layers because the strength of the excitatory currents is not sufficient to overcome the source/load mismatch caused by the coupling gradient. They also do not spread downwards since cells below were uncoupled. As mean automaticity $<\alpha>$ increases, so does the area to which the local waves spread. They then start to interact with each other, forming spirals and other dynamic patterns (Fig.5. and the supplemental video file). The local waves remain contained within the boundary layer until $<\alpha>$ reaches the level of 0.117. Afterwards waves start to exit the boundary and spread into the upper layers. The changing pattern of spirals and ectopic sources continues to exist within the boundary layer, but these sources were interacting with the waves returning from the upper zone. The quenching effect of these returning waves, amplified by a strong coupling in the upper layers, can be seen in the Fig.5 video. Notably, if one observes the events from the top layer, the overall activity of the boundary would appear as individual, somewhat irregular, ectopic beats exiting at random places.

The readers are asked to view the supplemental video files as they are an essential part of this report. Indeed, it is not trivial to reflect the observed dynamic events using a few sample frames such as those shown in Fig. 5A. The time traces from two individual cells (one within the boundary and the other from the well-coupled, upper layer) recorded as a change in cell membrane potential or intracellular calcium (Fig.5C) confirm that the records of calcium transients essentially reproduce the transmembrane potential recordings.

We have run several simulations using different spatial distributions of cells with all other parameters were identical. For each case, when $<\alpha>$ values were substantially below critical, the local waves remained locked within layers of intermediate coupling due to the source/load mismatch. When $<\alpha>$ values exceeded $<\alpha>_{crit}$, the waves started to propagate into the upper layers. Each case of spatial distribution $\eta(x,y)$ gave somewhat different scenarios of the generation of local waves and their escape.

The boundary layer shown in Fig.5 has a very steep coupling gradient (D values ranging from $10^{-5}$ to $10^{-1}$ cm$^2$/sec) which occurs over a narrow, 50 cell-wide layer of cells. Therefore, when local waves fuse and escape, it leads to an immediate activation of the entire upper layer (seen as a single yellow flash, Fig.5A). This is because the upper layer is fully coupled and conduction velocity there is high, as compared to the physical size of the media represented by the box. To obtain more information about what affects the generation and escape of local waves we "zoomed in" our simulations to the range of *D* values occurring just below the interface between the boundary layer and a well-coupled state. Therefore, in the next set of studies (Figs.6-9), we considered events within a less steep coupling gradient (from $10^{-5}$ to $10^{-3}$ cm$^2$/sec) with the upper layer corresponding to a weaker coupled network (D=$10^{-3}$ cm$^2$/sec). A much smaller wavelength (defined as the product of action potential duration and conduction velocity) allows clear visualization of the drift and escape of the tips of individual spiral waves into the upper layer.

Impact of cell heterogeneity. Our previous studies have suggested that individual cell heterogeneity is required to generate local waves (17). Specifically, if all members of the cell network were identical (in other words, dispersion coefficient, *δ*, was set to zero), the network would be either quiescent or all cells would fire simultaneously. Thus we studied how the degree of automaticity heterogeneity affects the generation of local waves. As detailed in the Methods section, heterogeneity of automaticity is determined by the two coefficients: *δ* and *η(x, y)*. So, first we varied the dispersion coefficient, *δ*, while spatial distribution *η(x, y)* was kept the same. In other words, the location of a cell with the highest *α* was the same, thereby the initial ectopic source was in the same place (see Fig. 6 and the corresponding video file). The overall result was when the degree of dispersion was low the ectopic activity started later and in fewer places.

When δ was kept the same, but the spatial distribution of cells, *η(x,y)* was varied, the location of individual ectopic sources was altered. The overall result, however, remained essentially the same, i.e. the effect of the specific location of individual cells within a static boundary was minor (data not shown). The spatial distribution of cells played a larger role when the boundary became dynamic, i.e., when it started to move in space as detailed immediately below.

Moving Boundary. Our next step was to fix the mean automaticity <α> at a value associated with ectopic activity and to investigate the effect of boundary movement (Fig.2. step 4). With a static spatial distribution, an ectopic source with the shortest period became dominant, creating a steady state pattern of events. In contrast, when the boundary was moved, the dominant role passed from one ectopic source to another, such that the system was constantly in a transient state. This increased the possibility of wavebreaks being formed and escaping into the better coupled layer (Fig.7).

Fig.7 also illustrates the role of the spatial distribution of cells, η(x,y). It shows frames from a protocol implemented for three different cell networks with all parameters identical, but with different spatial distributions. One can see that the escaping wavebreaks are formed at different locations and instants of time, leading to a different scenario in each case.

When the boundary moves, the spiral tip might exhibit a pin-unpin type of drift. This is a result of a combination of the microscopic cell heterogeneity and the macroscopic gradient of coupling. The direction of the drift observed numerically was in agreement with our experimental data (Fig.4A). The rightward drift of a counterclockwise spiral can be seen in Fig.8A and its corresponding video file. Another example can be seen in the video file corresponding to Fig.9A (right panel). An example of leftward drift of a clockwise spiral is shown in Fig.8B.

Drift of spiral waves caused by spatial gradients is well known in the theory of excitable media. The dominant mechanism of drift in our case was described in (30): the tip turns faster in the upper more coupled layer than in the lower, less coupled one. Therefore, the upper segments of the tip trajectory are shorter than the lower segments and the spiral drifts. However, the above mentioned theoretical works were for the systems with macroscopic gradients. In our case, we have microscopic heterogeneity, expressed via dispersion in automaticity values of individual cells. The effect of this heterogeneity is a pin-unpin type of drift, which was also seen in experiments (Fig.4A).

Let's consider the simulation shown in Fig.8A in a more detail. It is a numerical example of an "exaggerated" Pertsov-Ermakova drift mechanism. Visual analysis of this episode, as well as others, illustrates the general tendency for the tip to stick to clusters of cells with different properties, such as suppressed (Fig.8C) or elevated (Fig. 8D) automaticity. Notably, in the experiments, many, but not all, sites where spirals were pinned later became foci, i.e. they have underlying high automaticity clusters (Fig.4B). Of course, if it would have been a low-automaticity cluster, then cells will not spontaneously activate and such a cluster would be unnoticed (note: a cluster consisting of few cells would not impact macroscopic conduction to a measurable extend). As the gradient approaches, the spiral unpins and drifts. With the orientation of the tip as at $t$ =16.5 sec (Fig.8A, middle frame) and with the coupling above much higher than below, movement of the wave tip upwards is impeded by the source/load mismatch. The latter effectively reduces the automaticity at the tip and prevents it from turning. Hence we observe an almost straight segment of the tip trajectory as the tip "glides" along the coupling interface. Such straight motion continues until the tip reaches a more excitable locus and/or an area with a smaller coupling gradient, where it stops.

When the boundary moved at a higher speed, the probability of a spiral tip escaping into a more coupled area increased. As an example, Fig.9 shows parallel frames from the two simulations with all parameters identical, including spatial distribution $\eta(x,y)$ and dispersion coefficient, $\delta$. The only difference was the speed at which the boundary layer was moving. The fast moving boundary (7 cells/sec) caused the tip of a spiral wave to escape while the slower one (3.5 cells/sec) did not. Similar results were observed when we expanded our studies to three-dimensions (data not shown), a subject to be discussed in a separate publication.

DISCUSSION

Our data represents one of the first attempts to mimic complex boundary behavior in a heterogeneous cell network. Certain assumptions and simplifications had to be made to accomplish this task. These included the following:

*Experimental model.* Our study dealt with neonatal cardiomyocyte preparations which beat spontaneously albeit at slow rates (0.2-0.5 Hz at room temperature). Therefore, our interventions with either isoproterenol or barium simply elevated the endogenous automaticity of these cells. How can these events be related to myocardial tissue composed of quiescent adult ventricular myocytes?

The answer is our growing understanding that "true automaticity" and "triggered activity" are two sides of the same coin. Originally, the first meant a spontaneous, stand-alone pacemaker-like behavior, while the second required a preceding action potential to occur and was facilitated by calcium overload (31). However, the distinction between these two terms has become less and less clear, as "spontaneous" triggered activity due to leakiness of SR, calcium overload, elevated $I_{ns,Ca}$ and $I_{Cl(Na)}$, or upregulation of Na/Ca exchanger became apparent (32). Moreover, recent evidence suggests the main mechanism behind triggered activity in ventricular cells (calcium leak from the sarcoplasmic reticulum, followed by Ca/Na exchanger current) is also a major cause of automaticity in classical pacemakers, i.e. sinoatrial cells (33,34). Therefore, instead of opposing the concepts of automaticity and triggered activity, we refer to spontaneously active myocytes, regardless of their underlying mechanism, as "ectopics". The above arguments support the use of neonatal cardiomyocyte cultures as a model system to analyze behavior of ventricular tissue in which myocytes are made spontaneously active by pathological conditions.

*Wave monitoring*. Our experimental data was based on monitoring calcium transients using the calcium sensitive fluorescent indicator Fluo-4. This approach is widely used to follow propagating waves in cardiac muscle (26,35,36). Under control conditions, $Ca_{in}$ transients immediately follow electrical activity, and wave propagation patterns are essentially identical (37,38). Notably, during the initial stages of ectopic wave generation this sequence may be reversed, i.e. depolarization may follow the elevation of cytosolic calcium (39). Thus, in addition to its high fidelity, monitoring $Ca_{in}$ instead of transmembrane voltage insures that the earliest signs of ectopic activity are recorded. It is important to stress that intercellular waves observed in our studies represent spreading electrical activity and should not be confused with so-called intracellular calcium waves. Intracellular calcium waves are confined to individual myocytes and are much slower (0.1 mm/sec) than the velocities of local waves observed in our preparations. The latter ranged from 10-15 cm/sec in control conditions to 0.5 cm/sec in areas starting to recover from heptanol.

*Experimental means to increase cell automaticity*. In vivo, a multitude of factors, including extracellular potassium, pH, ATP, neuropeptides, and intra and extracellular calcium can cause a myocyte to reach depolarization threshold and fire an action potential (31). The subject of our studies, however, is not an individual cell, but a cell network. For the latter, the behavior is a result of mutual interaction, as well as timing and location of the individual myocytes *after* they passed their individual thresholds. Therefore, we simplified the multiple factors noted above as a single intervention that increased myocyte activity. The experimental data presented here corresponds to experiments with the beta-adrenergic agonist isoproterenol (16,40). An alternative way to increase myocyte automaticity is to use barium chloride, which inhibits inward potassium rectifier current ($I_{K1}$). The latter increases cardiomyocyte firing rate in a concentration-dependant manner (17). Our experiments with barium chloride produced phenomenologically similar results (data not shown).

*Numerical means of increasing automaticity*. Inhibition of Ik1 is not exactly a physiological way to mimic triggered activity during either ischemia or reperfusion. We used this approach as a convenient and established numerical tool to make cells automatic using Beeler-Reuter numerics. Much more detailed numerical models are required to fully simulate the effect of catecholamines or calcium-overload which induces triggered activity within individual myocytes, since one needs to take into account effects of calcium release from intracellular stores as well as geometry/spatial arrangements of adjacent intercellular compartments. The effect achieved by $I_{K1}$ inhibition is rather generic: it tilts the balance between inward and outward currents. The advantage of this particular model was the possibility of using the knowledge of the parameter space of the model, achieved in our previous work; there the model was carefully fitted to reproduce our experimental preparations (17). Further studies will be required to extend our conclusions derived from model systems to more relevant physiological scenarios. The semi-phenomenological description presented here simply suggests plausible scenarios and formulates interesting questions for subsequent more detailed studies.

*Conclusions*. With the assumptions and limitations noted above, the following conclusions were made from the bulk of the in vitro and numerical data. First, the data suggests that the combination of the two gradients (i.e., the spatial gradient in cell-to-cell coupling and the temporal gradient in cell automaticity) ensures that somewhere within the boundary there is a region where multiple ectopic sources are continuously being formed. They are highly localized focal points of activity, with activation spreading only to a few surrounding cells. The number of ectopic sources and the specific window of conditions when they occur are affected by the degree of the network heterogeneity. Secondly, our data argue that if the ectopically active layer is sufficiently wide and/or the overall cell automaticity rises, ectopic sources develop into target-like waves. If a coupling gradient and automaticity levels remain spatiotemporally fixed, the pattern of target-like sources persists and no spiral activity is observed. However, when cell automaticity rises and/or the boundary moves in space, the propagation patterns become non-stationary. This leads to multiple wavebreaks and spiral activity. Spiral waves typically demonstrate start-stop drifting behavior, as a result of competing forces between pinning force due to local heterogeneity and gradient-induced directional drift. The likelihood of spiral escape into better coupled tissue depends on the speed at which the boundary moves in space.

We are a long way from concluding that the patterns observed here can be found within inner layers of diseased myocardium and/or are the culprits of ectopic beats. Indeed, it is unlikely that in vivo the formation and escape of local waves occurs in *the exact way* it is portrayed in our figures. However, one has to consider that a wide window of conditions does occur when blood flows into the complex fractal surface of a previously occluded coronary artery bed. On such a moving boundary one can imagine a sharp coupling gradient together with rapidly recovering cell excitability as interstitial pH, potassium and oxygen levels are being restored while levels of interstitial norepinephrine are elevated. Therefore, it is hard to deny the probability that within small regions of moving boundary there will be a range of coupling and automaticity values which could breed local waves. Fig.10 places them in a larger pathophysiological context. The upper part explains how an individual, quiescent myocyte turns into a cell that spontaneously fires an action potential (41). How likely is it for spontaneously triggered activity to occur *synchronously* in a large number of cells? This is an important question because if activation does not involve a critical number of cells, a wave will not form. Therefore, an additional step is required between the triggered activity of individual cells and the arrhythmia on the level of the whole heart. The slow, local intercellular waves, similar to the regimes considered in this paper, appear to be the required step for the ectopic beats to be formed. We feel that the characterization of these regimes is important and that an awareness of such activity can facilitate its future detection in vivo.

Acknowledgments

We thank Drs. Matthew Kay, Alan Pumir and Valentin Krinsky for helpful discussions and Luther Swift for excellent technical support. Financial support of NIH (HL076722) and Engineering and Physical Sciences Research Council, UK (GR/S75314/01 and EP/S016391/1) is gratefully acknowledged.

Table 1. Numerical parameters used in individual simulations, corresponding figures & video supplements.

| Figure # & corresponding video file | Number of cells on x axis | Number of cells on y axis | Starting y coordinate which marks the lower part of the coupling gradient ($y_{1,\ t=0}$) | Starting y coordinate which marks the upper part of the coupling gradient ($y_{2,\ t=0}$) | Rate of boundary movement (cells/sec) | Coefficient D of the uncoupled layer ($cm^2/sec$) | Coefficient D of the well-coupled layer ($cm^2/sec$) | Starting automaticity value $<\alpha>_{t=0}$ | Rate of automaticity increase $d<\alpha>/dt$ ($sec^{-1}$) | Dispersion coefficient, $\delta$ |
|---|---|---|---|---|---|---|---|---|---|---|
| 5 | 100 | 125 | 25 | 75 | 0 | $10^{-5}$ | $10^{-1}$ | 0.08 | $10^{-3}$ | 0.5 |
| 6A | 100 | 100 | 25 | 50 | 0 | $10^{-5}$ | $10^{-3}$ | 0 | $10^{-3}$ | 0.25 |
| 6B | 100 | 100 | 25 | 50 | 0 | $10^{-5}$ | $10^{-3}$ | 0 | $10^{-3}$ | 0.5 |
| 7A,B,C* | 100 | 100 | 25 | 50 | 1/6 | $10^{-5}$ | $10^{-3}$ | 0.10 | 0 | 0.5 |
| 8D** | 100 | 100 | 65 | 95 | 1/10 | $5 \times 10^{-5}$ | $2 \times 10^{-3}$ | 0.12 | 0 | 0.5 |
| 9A | 100 | 100 | 70 | 95 | 3.5 | $10^{-5}$ | $10^{-3}$ | 0.12 | 0 | 0.5 |
| 9B | 100 | 100 | 70 | 95 | 7 | $10^{-5}$ | $10^{-3}$ | 0.12 | 0 | 0.5 |

*Fig.7A,B,C shows three simulations with identical parameters but different spatial cell distributions. Fig.8A&B uses the snapshots from the simulations shown in Fig.7A&C respectively. Fig. 8C is a magnified fragment of a data shown in Fig 8A.

** Fig 8D shows only a magnified fragment of the simulation data.

FIGURE LEGENDS

Fig.1: Experimental setup.

A. Schematics showing experimental setup which include multisyringe pump, custom made perfusion chamber and confocal imaging system in an inverted microscope configuration.

B. Cartoon illustrating experimental protocol which included the washout of heptanol (grey) by the isoproterenol-containing solution (white). Lighter shades of gray indicate the areas with a sharp coupling gradient.

Fig.2: Schematics showing the major steps involved in modeling the experimentally observed behavior. Details in Methods section.

Fig.3. Experimental data illustrating a continuous generation of target-like and spiral waves from the boundary layer.

A: Sequential images taken during one continuous recording. Each row illustrates the development of a particular wave pattern, including the appearance of multiple ectopic sources (first row), a clockwise rotating spiral (second row) and a single ectopic wave (third row). Arrows: direction of the wave spread. The dotted line shows the boundary of the uncoupled tissue. The acquisition window (black box) was moved during the experiment alongside the boundary to center each event. Note: because the washout process was much longer (> 2 min) than it took for an ectopic wave to spread through the entire field of view (< 1 sec) the boundary may appear static in these frames. The movement of the boundary can be clearly seen in the supplemental video file. Scale bar is 0.5 mm

B. Shape flattening of the spiral near the boundary. This occurs due to a progressive wavelength shortening as conduction velocity sharply dropped near the uncoupled zone. This also created a "shedding" effect, when wavefronts from two or three previous spiral rotations are seen within the thin layer of cells near the uncoupled area (arrows). Top row: four sequential frames. Bottom row: difference frames (pixel value in the frame above minus the pixel value in the preceding frame), showing spread of activation. The full sequence of events can be seen in the corresponding video file.

Fig.4. Drift and pinning of spiral waves in experimental settings.

A: The top panel illustrates the trajectory of a clockwise spiral. reconstructed from sequential confocal frames. Each dot indicates coordinates of the spiral tip within an individual frame. The spiral lasted 22 sec and drifted leftwards. It was pinned in three spots. An example of a counterclockwise spiral is shown below. It lasted for a shorter period (5 sec) and exhibited rightward drift. Full sequence of the events can be seen in the corresponding video file.

B. Illustration of an attachment of the spiral to an area of elevated automaticity. Top row: the last rotation of a spiral; tip position is marked by a red dot. Second row: target-like ectopic source appeared from the same spot immediately after spiral self-terminated.

Fig.5. Numerical studies: local waves forming within the boundary upon implementation of the first three steps shown in Fig.2.

A. Selected activation patterns from a simulation with growing $<\alpha>$. Before $<\alpha>$ reaches a certain value no activity is present. Then multiple spontaneous ectopic sources (t=26-29s) and local waves (t=29-36) start to appear within boundary layer, and at about t=37 sec this activity starts penetrating into a the well coupled layer. Full sequence of the events can be seen in the corresponding video file.

B. The average activity from the areas "A" and "B" (voltage recordings). The dotted lines point to the corresponding snapshots in the above panel A.

C. A comparison between action potential recordings and calcium transient recordings.

Fig.6. The impact of automaticity dispersion.

The mean automaticity $<\alpha>$ steadily grows with a rate of 0.001 sec$^{-1}$. The cartoon on the right side shows the automaticity dispersion along an arbitrary vertical line. *Top row:* Dispersion coefficient is $\delta = 0.25$. The number of ectopic sources is small and the activity appears later (t ~122 sec). *Bottom row*: Dispersion coefficient is $\delta = 0.5$. More ectopic sources are present and the activity appears sooner (t ~ 104 sec).

Fig.7. Impact of spatial cell distribution.

The mean automaticity $<\alpha>$ is fixed at 0.10 and the boundary moves downwards with the speed of 1/6 cells/sec. Cases shown in A, B and C differ only in the cells' spatial distribution $\eta(x,y)$.

Fig.8: Pin-unpin drift and the coupling gradient.

The mean automaticity $<\alpha>$ is fixed at 0.10 and the boundary moves downwards with the speed of 1/6 cells/sec. Selected frames from two different simulations illustrate the rightward (A) and leftward (B) drift alongside the boundary interface. The tip trajectory reflects an erratic pin-unpin behavior. Sequence shown in A can be seen in the corresponding video file. The two bottom panels show magnified pieces of tip trajectories on the background of the η(x,y) distribution in two different stimulations. The η(x,y) distributions have been smoothed by a 9×9 cell sliding window, and color-coding (blue component representing gk1) adjusted to embellish the pinning clusters (marked by stars). In C (fragment of the same simulation as in A), the spiral before unpinning was attached to a bright-blue (high gk1, suppressed automaticity) cluster. In D (a different simulation), the spiral before unpinning was attached to a dark-green (low gk1, elevated automaticity) cluster.

Fig.9: The speed of coupling gradient movement determines the probability of wavebreak escape.

The spatial distribution *η(x,y)* was identical for the cases shown in A and B. The mean automaticity $<\alpha>$ is fixed at 0.12. The only difference is the speed of the downward movement of the boundary.

A. Boundary moves slowly (3.5 cells/sec) and wavebreaks (i.e. tip of the spiral) do not escape. Therefore when boundary passes, no waves remain.

B. Boundary moves faster (7 cells/sec). The wavebreak escapes into the better-coupled layers. After the boundary passes, the spiral wave continues to rotate.

The full sequence of events shown in A&B can be seen in the corresponding video file.

Fig.10. Flow diagram which illustrates the potential pathophysiological significance of local waves.

The top part incorporates current beliefs (adapted from (41)) of how a variety of triggers might lead to spontaneous cell firing. RyR2: ryanodine receptors, CASQ2: calsequestrin, $I_{Cl(Ca)}$ calcium activated Cl current, $I_{ns(Ca)}$ nonspecific cation current. The local waves appear to be an essential intermediate step which facilitates the spread of the triggered activity from individual cells into well-coupled myocardial tissue.

DESCRIPTION OF SUPPLEMENTAL VIDEO FILES

SUPPLEMENT to Fig.3. (Quick Time Movie, 3.9Mb) Experimental video file illustrating the continuous generation of target-like (or ectopic) waves and spirals within the boundary layer. The acquisition window was moved during the experiment alongside the boundary to center each event. Propagation to the inner zone is impeded by the continuous presence of the uncoupler (heptanol).

SUPPLEMENT to Fig. 4 (Quick Time Movie, 3.9Mb) Experimental video file illustrating the distinct features of boundary spirals. An episode of the attachment of a spiral tip to an area of elevated excitability is shown. It manifests itself as an ectopic source emanating from the same area immediately after the spiral self-terminates. One can also see the shape flattening of the spiral near the boundary, due to a progressive wavelength shortening as conduction velocity sharply drops toward the inner zone. This creates an apparent "shedding" effect, when wavefronts from two or three previous spiral rotations are seen within the boundary layer.

SUPPLEMENT to Fig.5. (MPEG4, 5.3Mb) Numerical studies: formation of local waves within a boundary layer. Activation patterns developing for the boundary with growing mean automaticity $<\alpha>$ (see Fig. 5 legend for details). Color coding by components: red for transmembrane voltage, green for $D$ and and blue for $g_{K1}$.

SUPPLEMENT to Fig. 6 (MPEG4, 8.8Mb) Impact of cell heterogeneity. Mean automaticity $<\alpha>$ grows while the boundary is fixed in space. The coefficients of dispersion are different: $\delta = 0.5$ for the study shown on the left and $\delta = 0.25$ for the study shown on the right. The videos correspond to the events shown in Fig. 5A, specifically they show 100-180 sec of the simulation sequence.

SUPPLEMENT to Fig. 8 (MPEG4, 2.7Mb) Drift of a spiral along moving boundary. Drift with pinning and then escape into the well coupled zone. The boundary moves downwards slowly at a rate of 1/6 cell/sec. The video file corresponds to the events shown in Fig. 8A (first 25 sec of the simulation sequence).

SUPPLEMENT to Fig. 9 (MPEG4, 3.4Mb) Effect of the speed. Video on the left: boundary moves faster, spiral waves escape to the well coupled zone. Video on the right: boundary moves slower, spiral tip does not escape. In both simulations, the boundary starts moving at t=10 sec. The videos correspond to the events depicted in Fig.9, specifically they show the first 30 sec of the simulation sequence.

REFERENCES

1. Dhein S. Cardiac ischemia and uncoupling: gap junctions in ischemia and infarction. *Adv Cardiol*. 2006;42:198-212.
2. Stevenson WG, JN Weiss, I Wiener, K Nademanee. Slow conduction in the infarct scar: relevance to the occurrence, detection, and ablation of ventricular reentry circuits resulting from myocardial infarction. *Am Heart J*. 1989;117:452-67.

3. Peters NS. New insights into myocardial arrhythmogenesis: distribution of gap-junctional coupling in normal, ischaemic and hypertrophied human hearts. *Clin Sci (Lond)*. 1996;90:447-52.
4. Elizari MV, PA Chiale. Cardiac arrhythmias in Chagas' heart disease. *J Cardiovasc Electrophysiol*. 1993;4:596-608.
5. Kies P, M Bootsma, J Bax, MJ Schalij, EE van der Wall. Arrhythmogenic right ventricular dysplasia/cardiomyopathy: screening, diagnosis, and treatment. *Heart Rhythm*. 2006;3:225-34.
6. Peters NS, AL Wit. Myocardial architecture and ventricular arrhythmogenesis. *Circulation*. 1998;97:1746-54.
7. Mills WR, N Mal, MJ Kiedrowski, R Unger, F Forudi, ZB Popovic, MS Penn, KR Laurita. Stem cell therapy enhances electrical viability in myocardial infarction. *J Mol Cell Cardiol*. 2007;42:304-14.
8. White RL, JE Doeller, VK Verselis, BA Wittenberg. Gap junctional conductance between pairs of ventricular myocytes is modulated synergistically by H+ and Ca++. *J Gen Physiol*. 1990;95:1061-75.
9. Yamada KA, J McHowat, GX Yan, K Donahue, J Peirick, AG Kleber, PB Corr. Cellular uncoupling induced by accumulation of long-chain acylcarnitine during ischemia. *Circ Res*. 1994;74:83-95.
10. Lameris TW, S de Zeeuw, G Alberts, F Boomsma, DJ Duncker, PD Verdouw, AJ Veld, AH van Den Meiracker. Time course and mechanism of myocardial catecholamine release during transient ischemia in vivo. *Circulation*. 2000;101:2645-50.
11. Warner MR, PL Wisler, TD Hodges, AM Watanabe, DP Zipes. Mechanisms of denervation supersensitivity in regionally denervated canine hearts. *Am J Physiol*. 1993;264:H815-20.
12. Karmazyn M, XT Gan, RA Humphreys, H Yoshida, K Kusumoto. The myocardial Na(+)-H(+) exchange: structure, regulation, and its role in heart disease. *Circ Res*. 1999;85:777-86.
13. Zipes DP, M Rubart. Neural modulation of cardiac arrhythmias and sudden cardiac death. *Heart Rhythm*. 2006;3:108-13.
14. Arutunyan A, DR Webster, LM Swift, N Sarvazyan. Localized injury in cardiomyocyte network: a new experimental model of ischemia-reperfusion arrhythmias. *Am J Physiol Heart Circ Physiol*. 2001;280:H1905-15.
15. Arutunyan A, L Swift, N Sarvazyan. Multiple injury approach and its use for toxicity studies. *Cardiovasc Toxicol*. 2004;4:1-10.
16. Arutunyan A, A Pumir, VI Krinsky, LM Swift, N Sarvazyan. Behavior of ectopic surface: effects of beta-adrenergic stimulation and uncoupling. *Am J Physiol Heart Circ Physiol*. 2003.
17. Pumir A, A Arutunyan, V Krinsky, N Sarvazyan. Genesis of ectopic waves: role of coupling, automaticity, and heterogeneity. *Biophys J*. 2005;89:2332-49.
18. Beeler GW, H Reuter. Reconstruction of the action potential of ventricular myocardial fibres. *J Physiol*. 1977;268:177-210.
19. Pumir A, V Krinsky. Unpinning of a rotating wave in cardiac muscle by an electric field. *J Theor Biol*. 1999;199:311-9.
20. Keener J, J Sneyd. *Mathematical Physiology*: Springer Verlag; 1998.
21. Silva J, Y Rudy. Mechanism of pacemaking in I(K1)-downregulated myocytes. *Circ Res*. 2003;92:261-3.

22. Dhamoon AS, J Jalife. The inward rectifier current (IK1) controls cardiac excitability and is involved in arrhythmogenesis. *Heart Rhythm*. 2005;2:316-24.
23. Masuda H, N Sperelakis. Inwardly rectifying potassium current in rat fetal and neonatal ventricular cardiomyocytes. *Am J Physiol*. 1993;265:H1107-11.
24. Wahler GM. Developmental increases in the inwardly rectifying potassium current of rat ventricular myocytes. *Am J Physiol*. 1992;262:C1266-72.
25. Shen JB, M Vassalle. Barium-induced diastolic depolarization and controlling mechanisms in guinea pig ventricular muscle. *J Cardiovasc Pharmacol*. 1996;28:385-96.
26. Agladze K, MW Kay, V Krinsky, N Sarvazyan. Interaction between Spiral and Paced Waves in Cardiac Tissue. *Am J Physiol Heart Circ Physiol*. 2007.
27. Entcheva E, SN Lu, RH Troppman, V Sharma, L Tung. Contact fluorescence imaging of reentry in monolayers of cultured neonatal rat ventricular myocytes. *J Cardiovasc Electrophysiol*. 2000;11:665-76.
28. Iravanian S, Y Nabutovsky, CR Kong, S Saha, N Bursac, L Tung. Functional reentry in cultured monolayers of neonatal rat cardiac cells. *Am J Physiol Heart Circ Physiol*. 2003;285:H449-56.
29. Kleber AG, Y Rudy. Basic mechanisms of cardiac impulse propagation and associated arrhythmias. *Physiol Rev*. 2004;84:431-88.
30. Pertsov AM, EA Ermakova. Mechanism of the drift of a spiral wave in an inhomogeneous medium. *Biofizika*. 1988;33:338-342.
31. Carmeliet E. Cardiac ionic currents and acute ischemia: from channels to arrhythmias. *Physiol Rev*. 1999;79:917-1017.
32. Ter Keurs HE, PA Boyden. Calcium and arrhythmogenesis. *Physiol Rev*. 2007;87:457-506.
33. Bogdanov KY, VA Maltsev, TM Vinogradova, AE Lyashkov, HA Spurgeon, MD Stern, EG Lakatta. Membrane potential fluctuations resulting from submembrane Ca2+ releases in rabbit sinoatrial nodal cells impart an exponential phase to the late diastolic depolarization that controls their chronotropic state. *Circ Res*. 2006;99:979-87.
34. Bers DM. The beat goes on: diastolic noise that just won't quit. *Circ Res*. 2006;99:921-3.
35. Bub G, L Glass, NG Publicover, A Shrier. Bursting calcium rotors in cultured cardiac myocyte monolayers. *Proc Natl Acad Sci U S A*. 1998;95:10283-7.
36. Entcheva E, Y Kostov, E Tchernev, L Tung. Fluorescence imaging of electrical activity in cardiac cells using an all-solid-state system. *IEEE Trans Biomed Eng*. 2004;51:333-41.
37. Entcheva E, H Bien. Macroscopic optical mapping of excitation in cardiac cell networks with ultra-high spatiotemporal resolution. *Prog Biophys Mol Biol*. 2006;92:232-57.
38. Laurita KR, A Singal. Mapping action potentials and calcium transients simultaneously from the intact heart. *Am J Physiol Heart Circ Physiol*. 2001;280:H2053-60.
39. Lakkireddy V, G Bub, P Baweja, A Syed, M Boutjdir, N El-Sherif. The kinetics of spontaneous calcium oscillations and arrhythmogenesis in the in vivo heart during ischemia/reperfusion. *Heart Rhythm*. 2006;3:58-66.
40. Schlotthauer K, DM Bers. Sarcoplasmic reticulum Ca(2+) release causes myocyte depolarization. Underlying mechanism and threshold for triggered action potentials. *Circ Res*. 2000;87:774-80.
41. Rubart M, DP Zipes. Mechanisms of sudden cardiac death. *J Clin Invest*. 2005;115:2305-15.

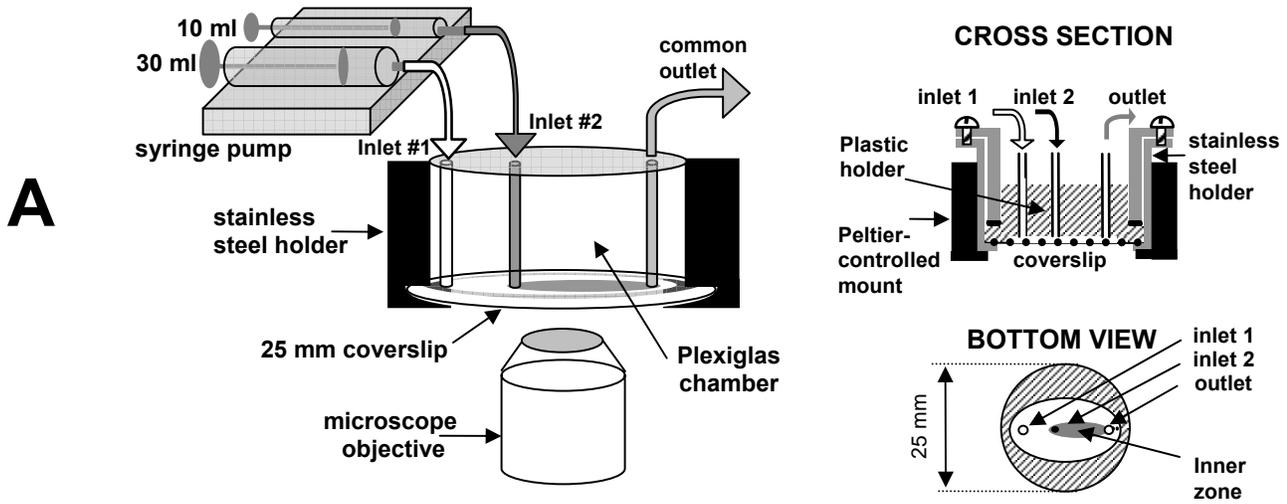

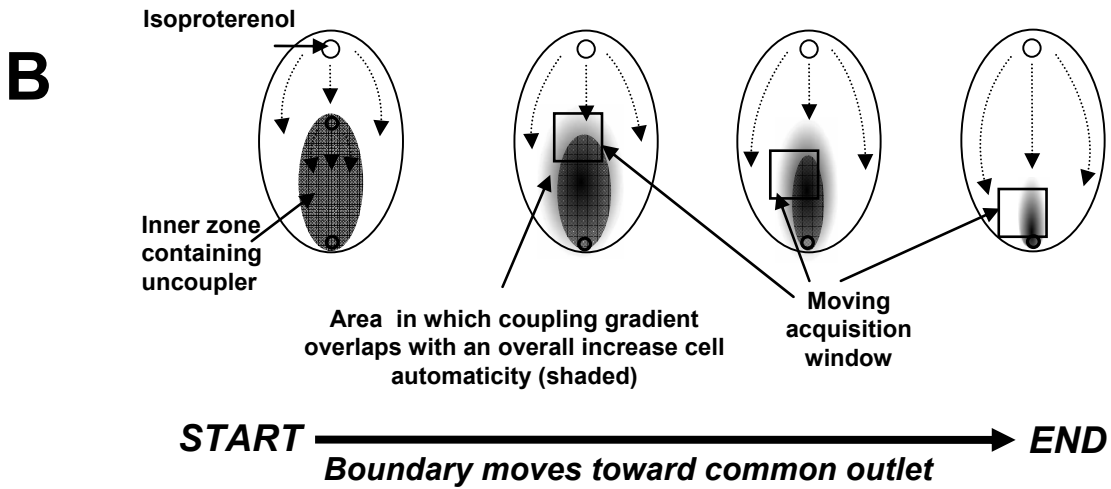

Fig.1. Cartoon of experimental protocol

**Step 1: Spatial gradient of coupling**  **Step 2: Spatial profile of automaticity, α**

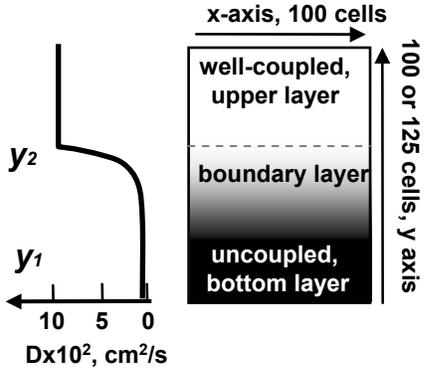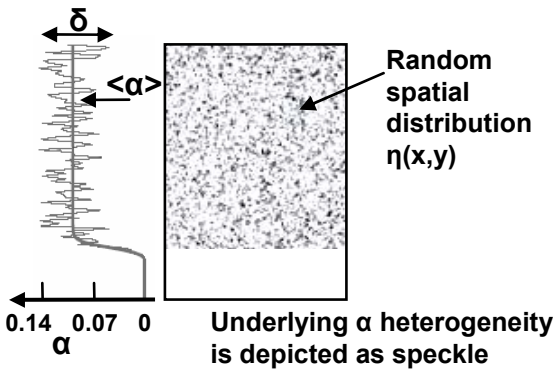

**Step 3: Increase in mean automaticity $\langle\alpha\rangle$**

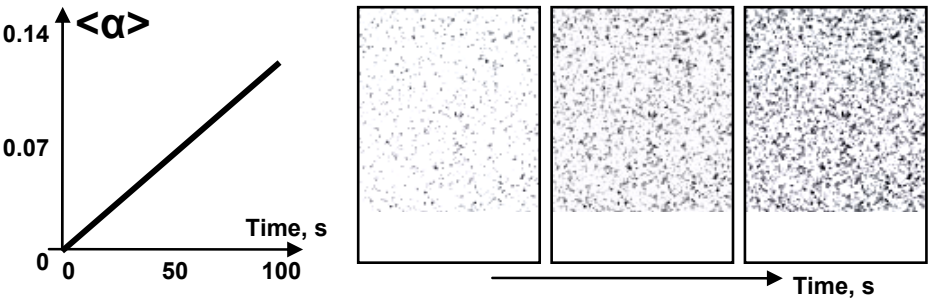

**Step 4: Boundary moves downwards.**

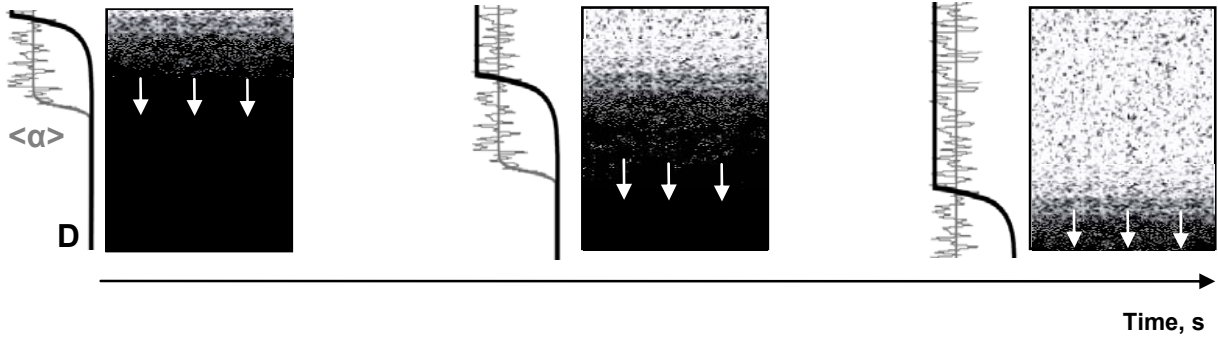

Fig.2. Cartoon of numerical protocol

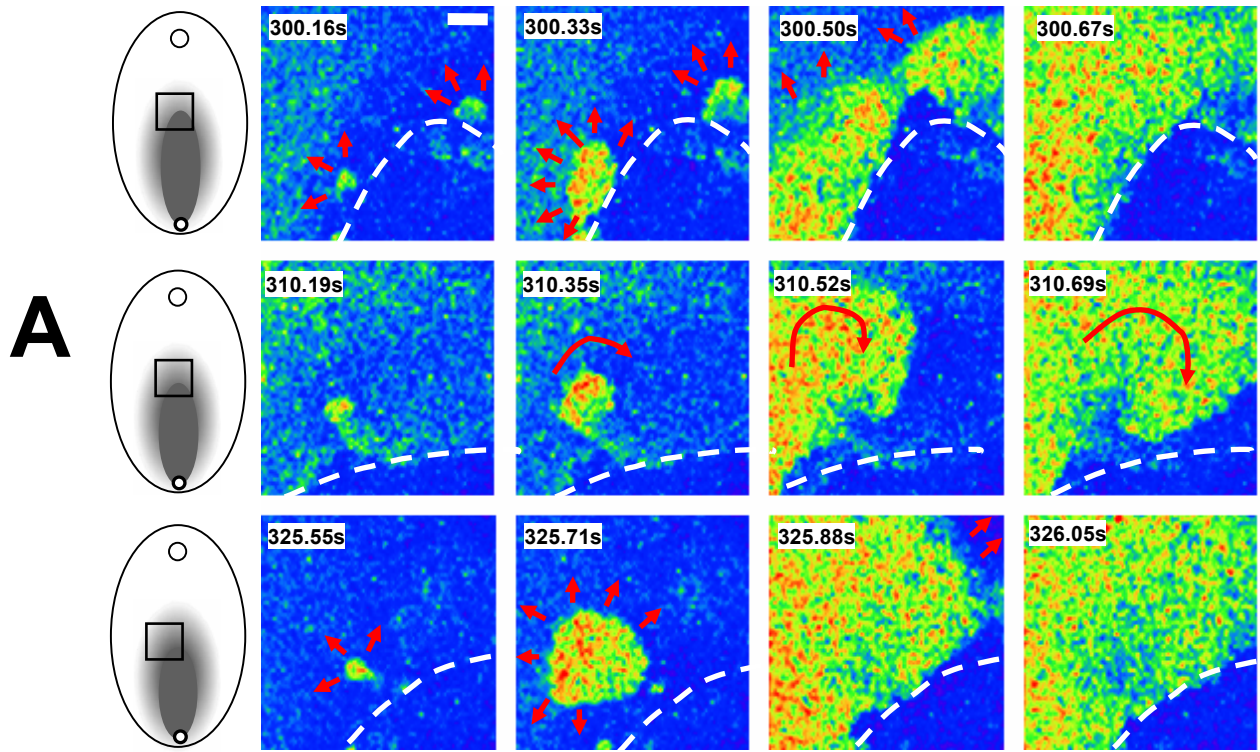

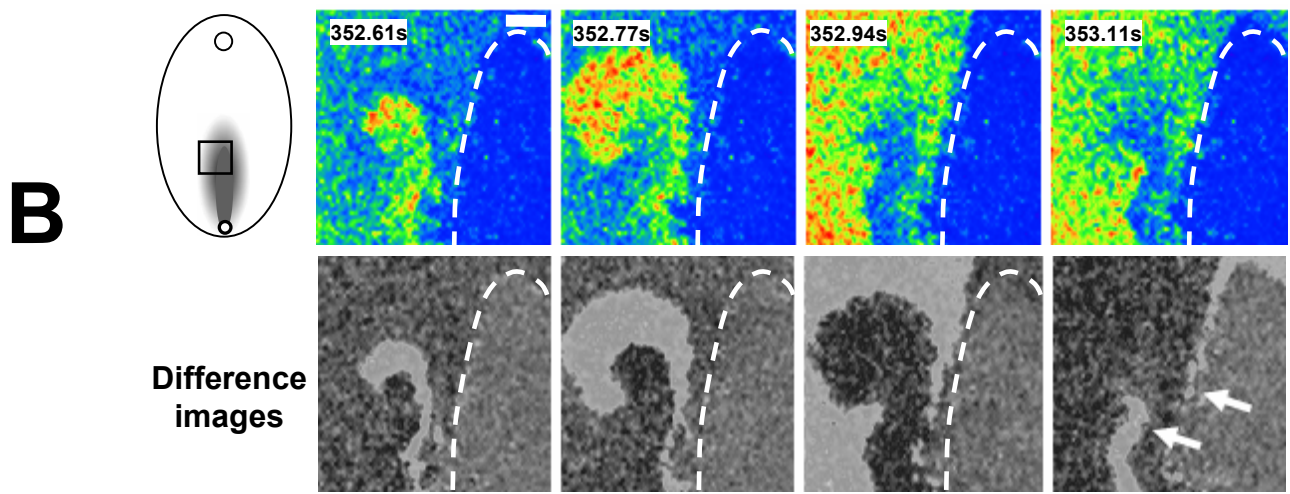

**Fig.3**

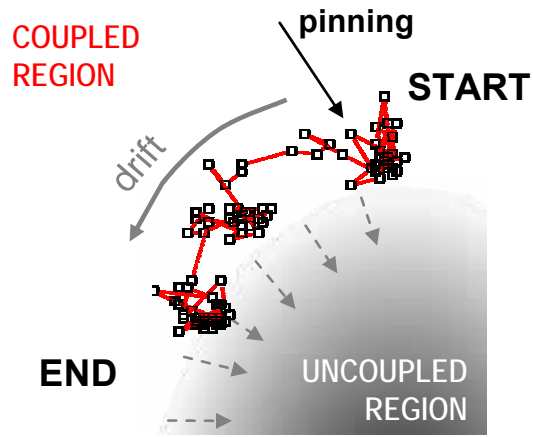
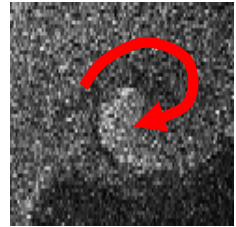

Duration: 22 sec
chirality: clockwise
Drift: leftward

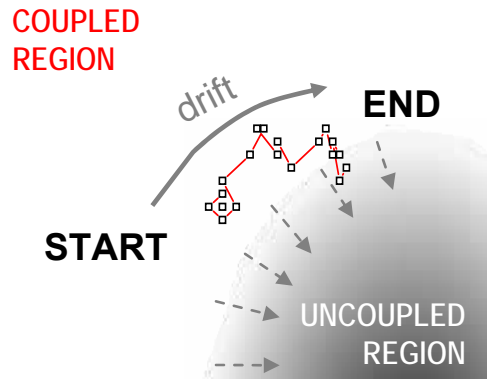
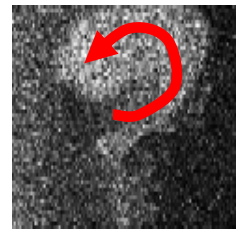

Duration: 5 sec
chirality: counterclockwise
Drift: rightward

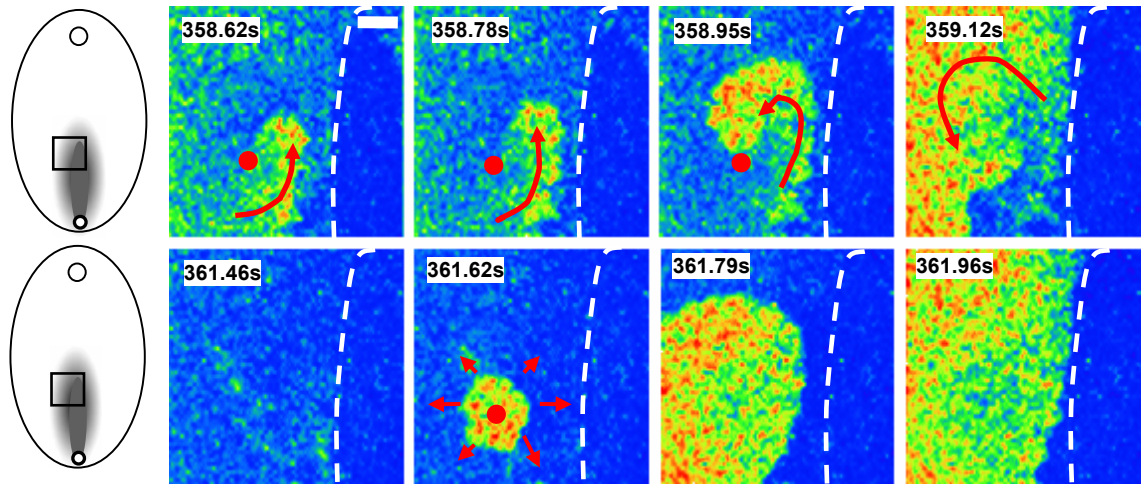

**Fig.4**

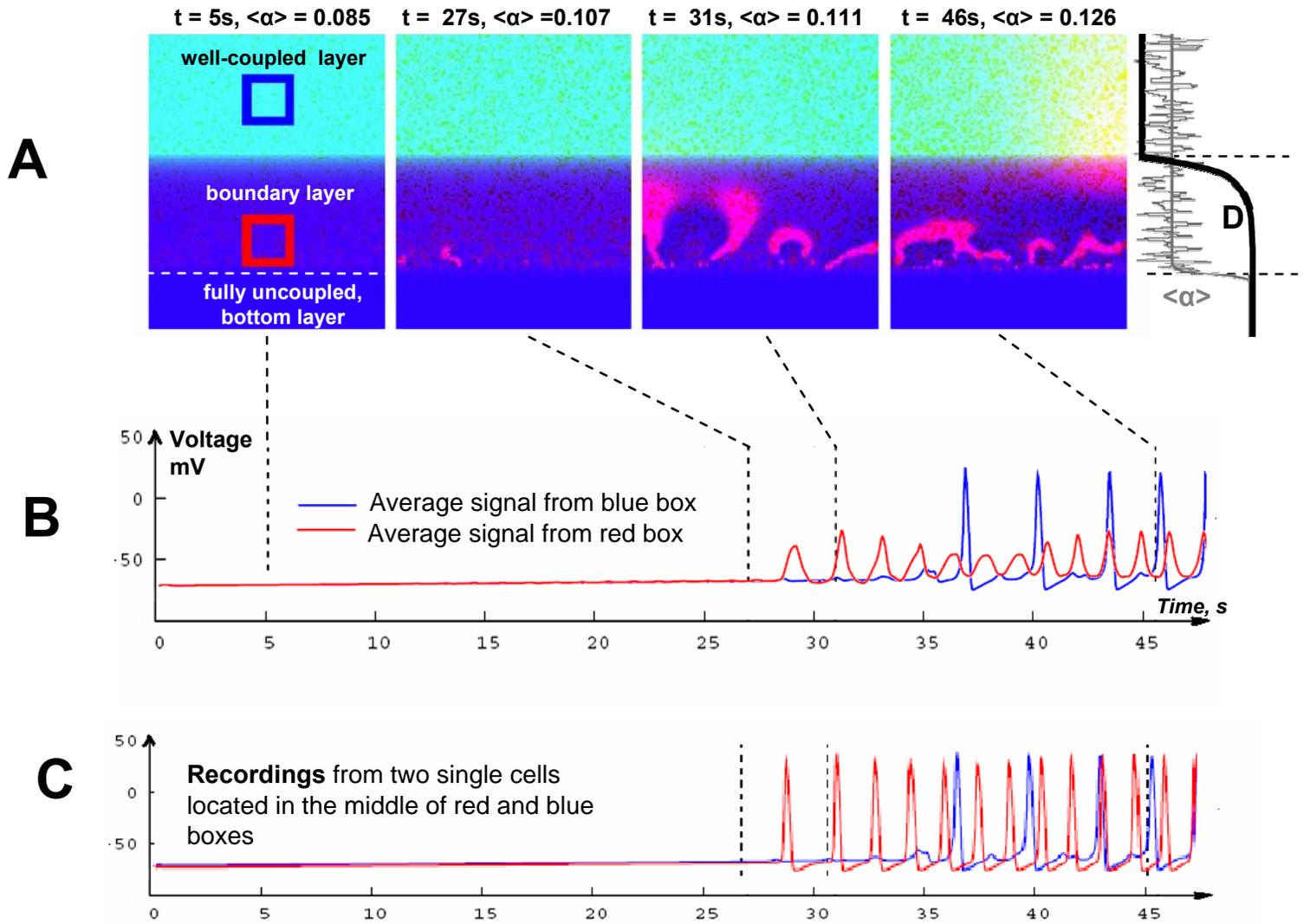

**Fig.5**

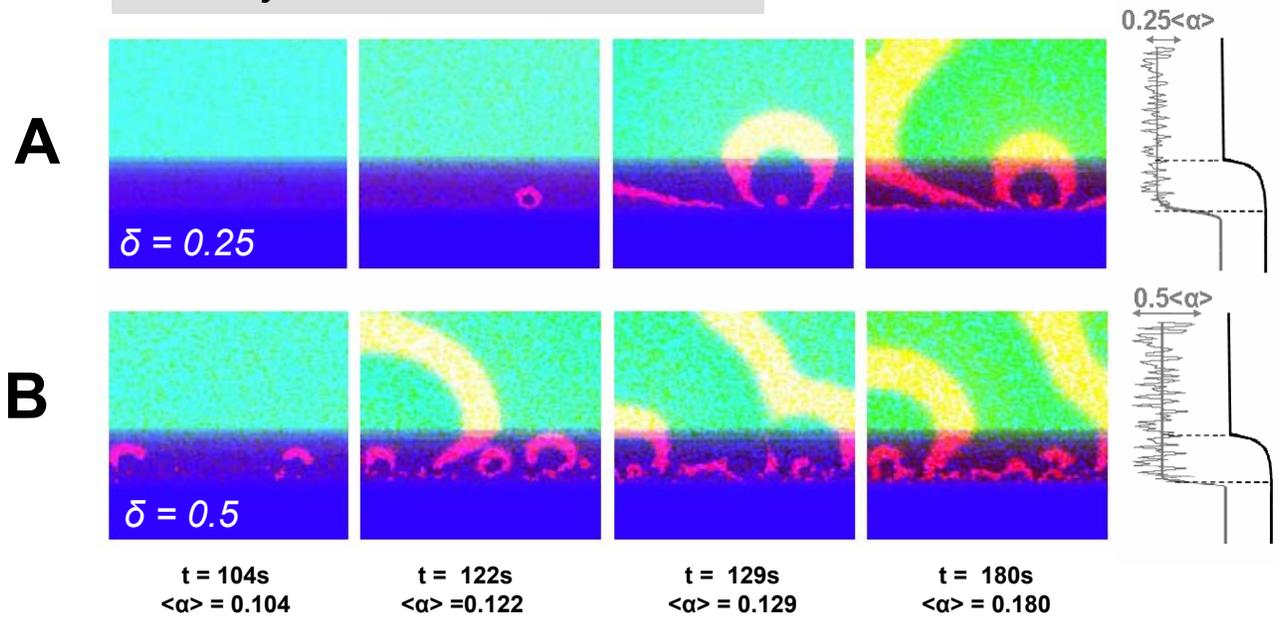

**Fig.6**

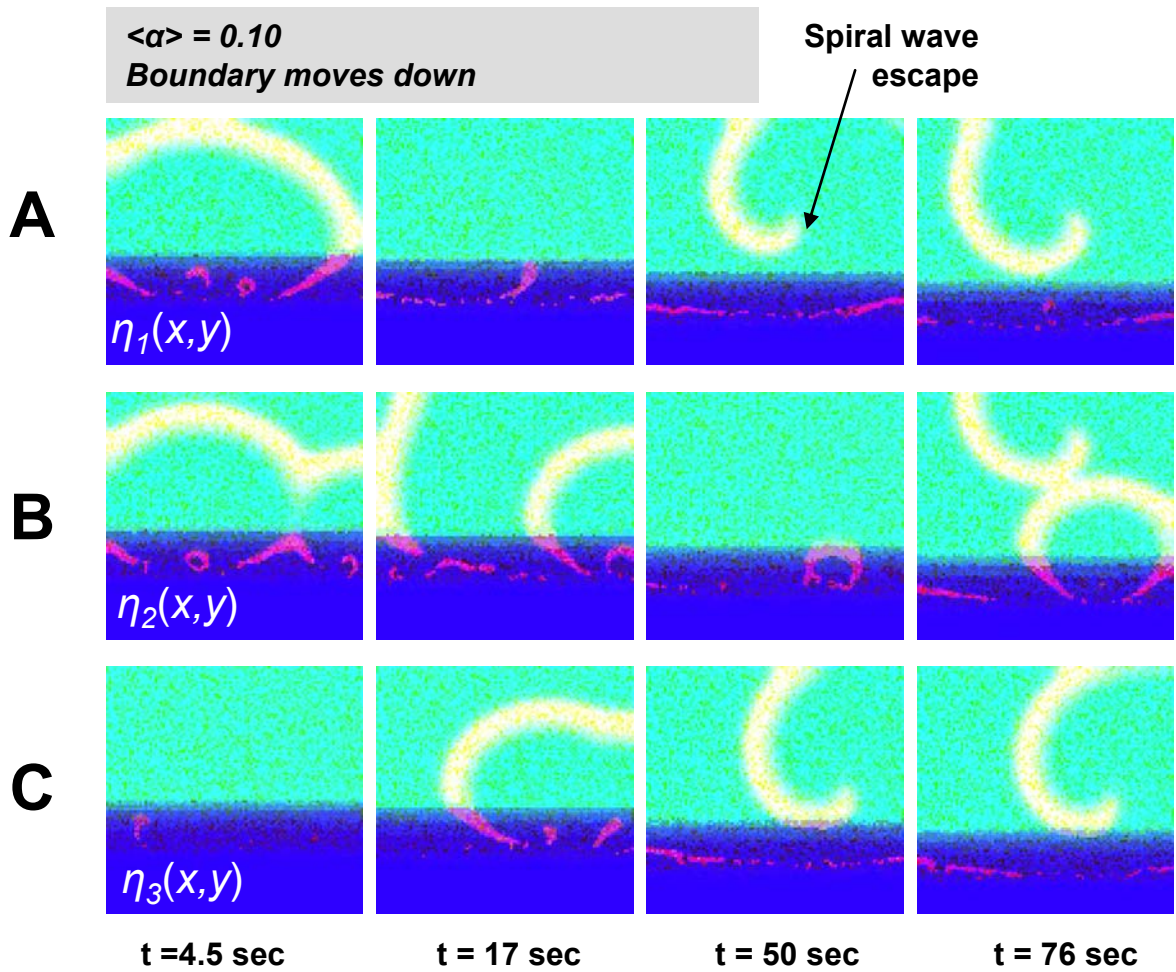

Fig.7

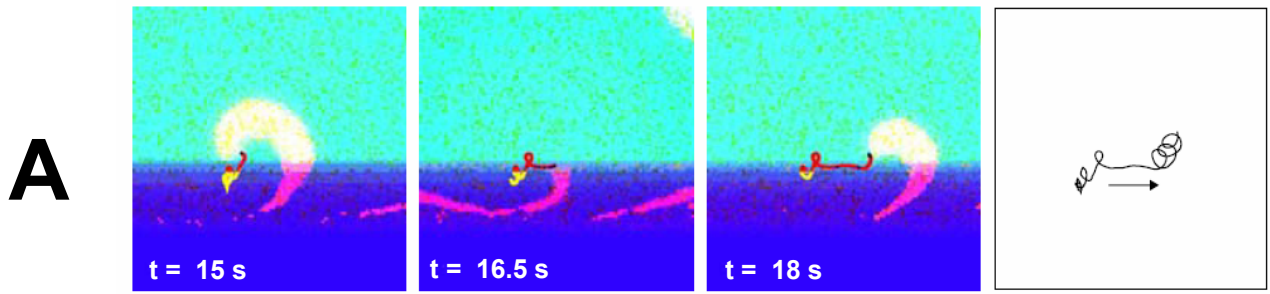

**Chirality: counterclockwise, spiral tip drift: rightward**

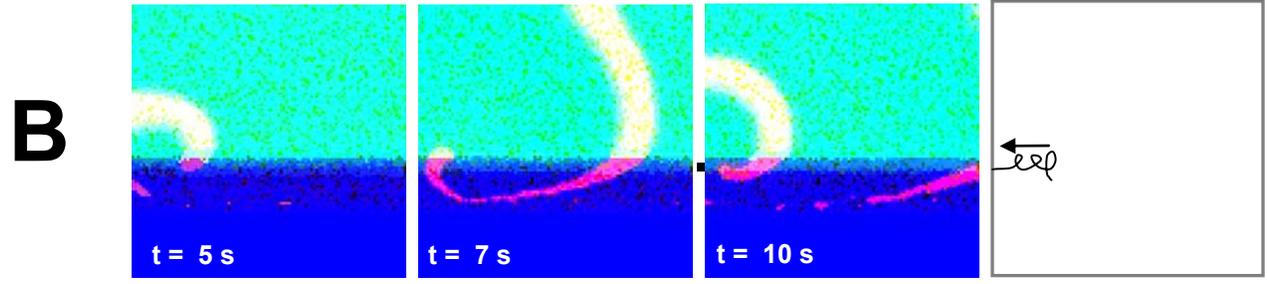

**Chirality: clockwise, spiral tip drift: leftward**

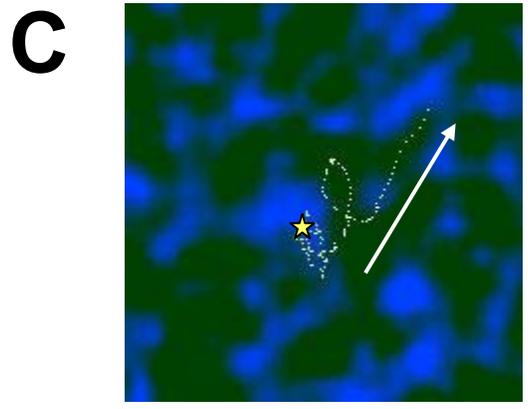

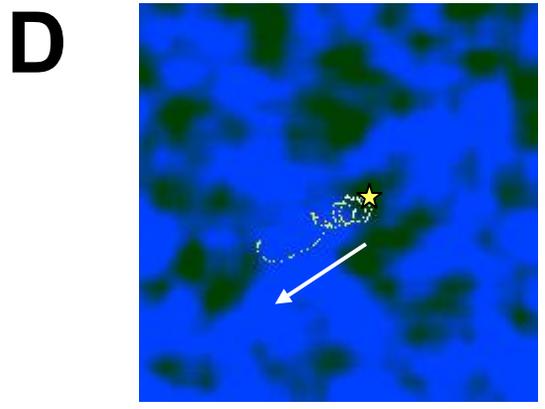

**Fig.8**

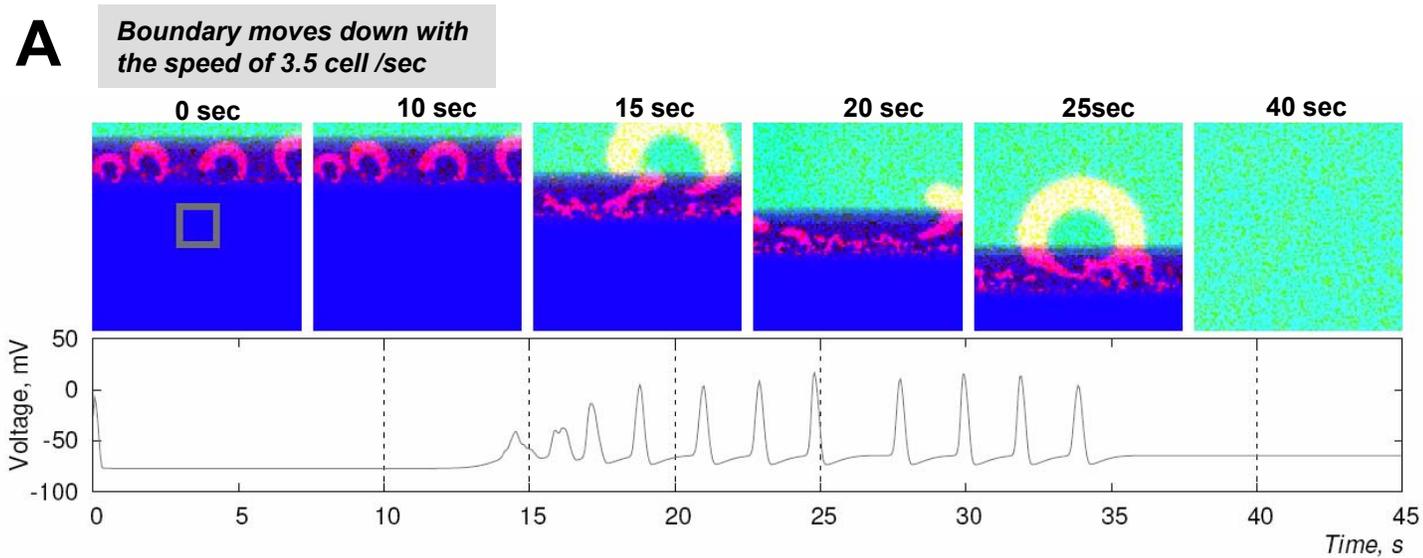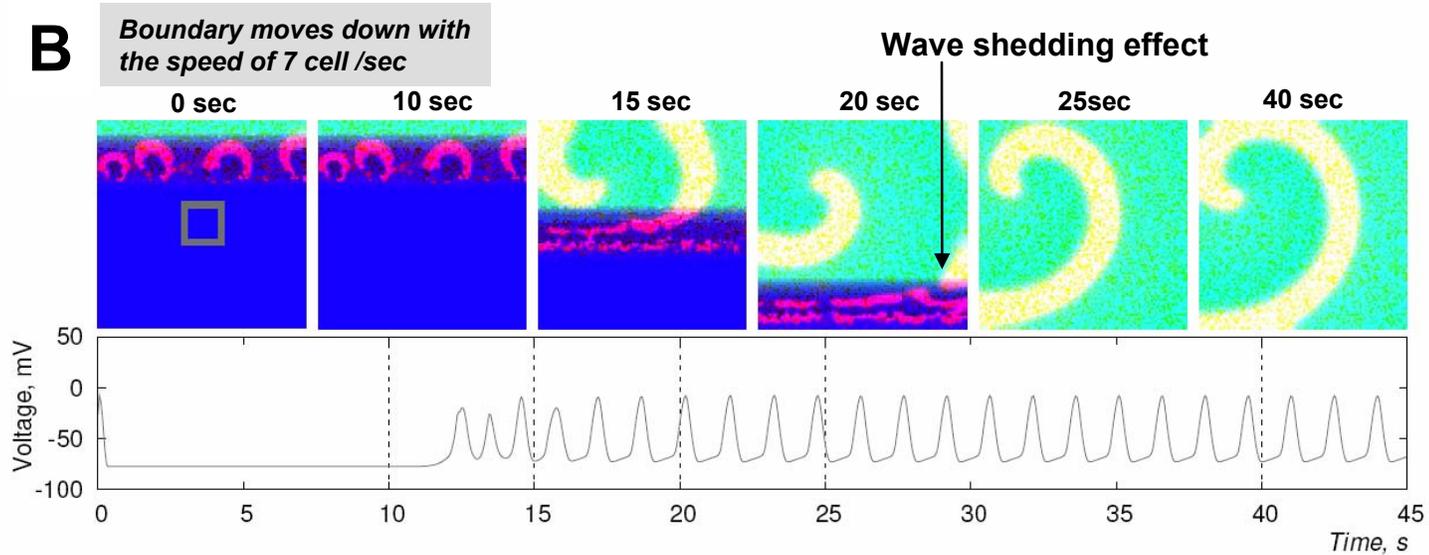

Fig.9

| | |
|---|---|
| **Myocyte**<br>Media size ~ 100 um<br>Velocity ~ 100 µm/s<br>Propagation:<br>Ca Induced-Ca<br>Release based | 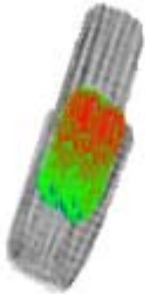 |
| **Neighboring cell network**<br>Media size ~ 1 mm<br>Velocity ~ 1-10 mm/s<br>Propagation:<br>Action potential based | 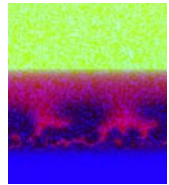 |
| **Myocardium**<br>Media size ~ 10 cm<br>Velocity ~ 50 cm/s<br>Propagation:<br>Action potential based | 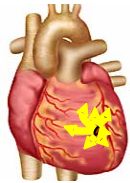 |

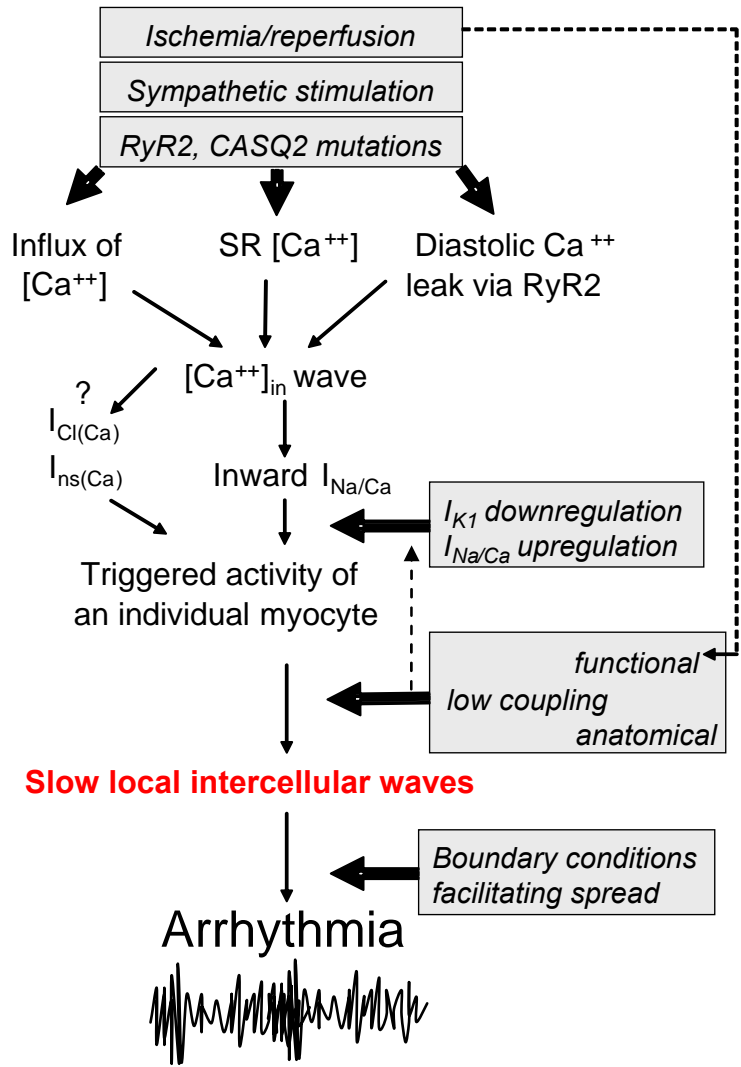

**Fig.10**